\documentclass[reprint,aps,prl,superscriptaddress,twocolumn,floatfix]{revtex4-1} 

\usepackage{graphicx}
\usepackage{amsmath}
\usepackage{bm}
\usepackage{color}
\usepackage{soul}
\usepackage{ulem}

\begin{document}

\title{Dissipative solitary wave at the interface of a binary complex plasma}

\author{Wei Sun}
\affiliation{College of Science, Donghua University, 201620 Shanghai, PR China}
\author{Mierk Schwabe}
\author{Hubertus M Thomas}
\affiliation{Institut f\"ur Materialphysik im Weltraum, Deutsches Zentrum f\"ur Luft- und Raumfahrt (DLR), 82234 We{\ss}ling, Germany}
\author{Andrey M Lipaev}
\author{Vladimir I Molotkov}
\author{Vladimir E Fortov}
\affiliation{Joint Institute for High Temperatures, 125412 Moscow, Russia}
\author{Yan Feng}
\affiliation{Center for Soft Condensed Matter Physics and Interdisciplinary Research, College of Physics, Optoelectronics and Energy, Soochow University, 215006 Suzhou, PR China}
\author{Yi-Fei Lin}
\author{Jing Zhang}
\affiliation{College of Science, Donghua University, 201620 Shanghai, PR China}
\author{Cheng-Ran Du}
\email{chengran.du@dhu.edu.cn}
\affiliation{College of Science, Donghua University, 201620 Shanghai, PR China}
\affiliation{Member of Magnetic Confinement Fusion Research Centre, Ministry of Education, PR China}

\begin{abstract}
The propagation of a dissipative solitary wave across an interface is studied in a binary complex plasma. The experiments were performed under microgravity conditions in the PK-3 Plus Laboratory on board the International Space Station using microparticles with diameters of 1.55~$\mu$m and 2.55~$\mu$m immersed in a low-temperature plasma. The solitary wave was excited at the edge of a particle-free region and propagated from the sub-cloud of small particles into that of big particles. The interfacial effect was observed by measuring the deceleration of particles in the wave crest. The results are compared with a Langevin dynamics simulation, where the waves were  excited by a gentle push on the edge of the sub-cloud of small particles. Reflection of the wave at the interface is induced by increasing the strength of the push. By tuning the ion drag force exerted on big particles in the simulation, the effective width of the interface is adjusted. We show that the strength of reflection increases with narrower interfaces.
\end{abstract}

\maketitle

{\it Introduction}
A complex plasma is a weakly ionized gas containing small solid particles \cite{Fortov:2005,Morfill:2009,Chaudhuri:2011}. The particles are highly charged by collecting ions and electrons, and interact with each other via screened Coulomb interaction. With video microscopy, localized structures and dynamics can be recorded in experiments so that various phenomena such as formation of crystal lattice \cite{chu:1994,thomas:1994,thomas:1996,zuzic:2006,williams:2012,nosenko:2009}, propagation of acoustic waves \cite{Rao:1990,Barkan:1995,Schwabe:2007,Merlino:2014,Thomas:2006,Menzel:2010,Williams:2014,Heinrich:2012,Teng:2009,Tsai:2016}, and the development of instabilities \cite{Schwabe:2009,Couedel:2010,Mikikian:2007,Rosenberg:1996,Heidemann:2011a} can be studied at the kinetic level. A binary complex plasma contains two differently sized types of microparticles. These particle types can either be mixed \cite{Du:2016} or form a phase separated system \cite{Wysocki:2010,Jiang:2011,Du:2012}. The later is usually caused by spinodal decomposition if certain requirement is met \cite{Ivlev:2009}. However, sometimes an imbalance of external forces can also lead to phase separation \cite{Killer:2016} even though the criteria of spinodal decomposition are not fulfilled. In both scenarios, an interface between separated phases emerges and enables us to study various interfacial phenomena.

In past years, the study of wave propagation at the interface has drawn much attention. The waves of interest include not only acoustic waves \cite{Wen:2009,Godin:2006} but also light waves \cite{Xiao:2010}, electromagnetic waves \cite{Bertin:2012}, spin waves \cite{Kajiwara:2010}, etc. Recently, it has been discovered that a ``collision zone'' and a ``merge zone'' exist close to the interface as self-excited waves propagate in binary complex plasmas \cite{Yang:2017}. However, for such continuous waves, the reflected wave from the interface (if it exists) is coupled to the next forward-propagating wave and thus cannot be resolved. Besides, due to the mechanism triggering the waves, namely two-stream instability \cite{Rosenberg:1996} coupled with heartbeat instability \cite{Heidemann:2011,Mikikian:2007}, the kinetic energy of microparticles is constantly exchanged with the plasma so that the wave amplitude may increase even in the presence of damping due to gas drag \cite{Epstein:1929}.

Unlike continuous waves, the propagation of externally excited solitary waves provides an excellent opportunity to study the fine features of the wave behavior at the interface. In granular matter, anomalous reflection was discovered at the interface of a chain of grains \cite{Nesterenko:2005}, and it was shown that the refraction follows Snell's law in a two-dimensional granular system \cite{Tichler:2013}. In complex plasmas, both compressive and dark solitary waves have been extensively studied experimentally and theoretically \cite{Bandyopadhyay:2008,Heidemann:2009,Samsonov:2002,Durniak:2010,Fortov:2005b}. Among those, particular attention has been paid to the head-on collision of two solitary waves \cite{Jaiswal:2016,Harvey:2010,Xue:2004,Ghosh:2011,Sharma:2014}.

In this letter, we study the propagation of a dissipative solitary wave across an interface in a binary complex plasma. The experiments were performed in the PK-3 Plus Laboratory on board the International Space Station. The solitary wave was excited at the edge of the central particle-free region called ''void`` and propagated from the sub-cloud of small particles into the that of big particles. The interfacial effect was observed by measuring the deceleration of particles in the wave crest. The results are compared with a Langevin dynamics simulation, where the waves were excited by a gentle push on the edge of the sub-cloud of small particles. By increasing the strength of the push, reflection at the interface is triggered. By tuning the ion drag force exerted on the big particles in the simulation, the width of the interface is adjusted. The dependence of the strength of reflection on the width of the interface is studied.

\begin{figure}[!ht]
\centerline{\includegraphics[bb=10 0 700 700, width=0.5\textwidth]{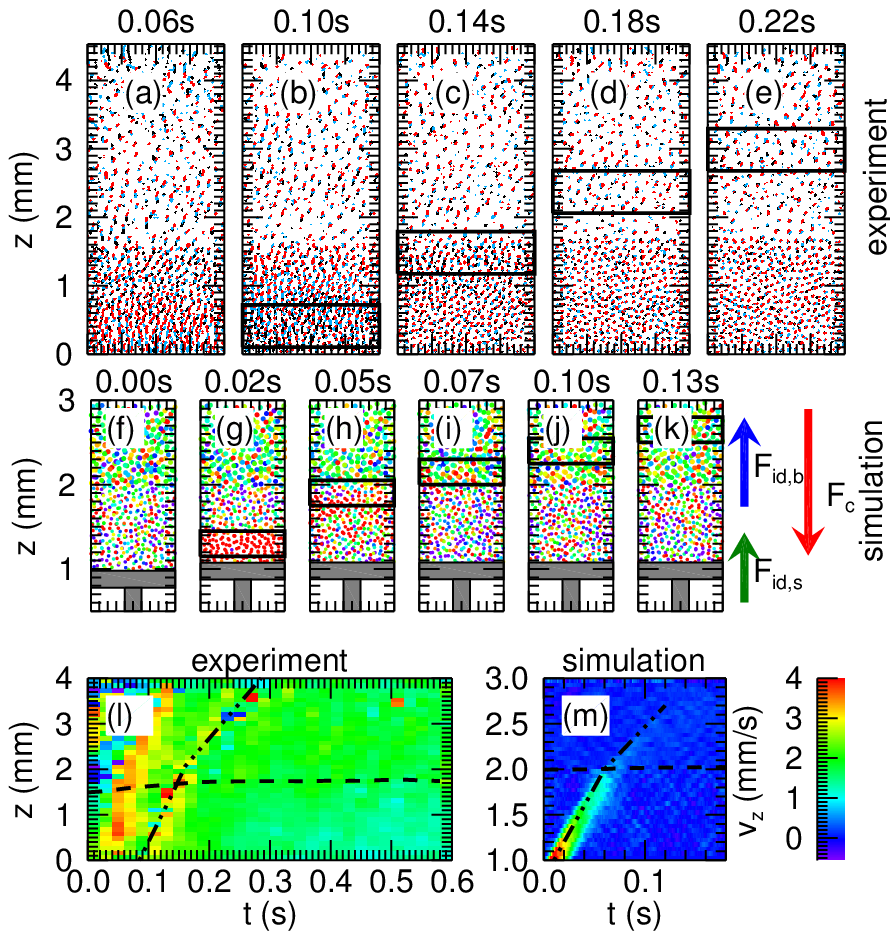}}
\caption{(Color online) Snapshots of a quasi-solitary wave in a binary complex plasma in an experiment (a-e) and in Langevin dynamics simulation (f-k). For the experiment, three consecutive images are overlaid and the two types of particles can be distinguished by the interparticle distance. (See supplemental material for the whole process.) For the simulation, the two types of particles can be distinguished by the particle size, and the color coded from violet to red shows the vertical velocity $v_z$. The evolution of the solitary wave in the experiment (l) and simulation (m). The interface is marked by a dashed curve and the wave crest is highlighted by a dashed-dotted curve. The color bar is truncated at $4$~mm/s.
\label{figure1}}
\end{figure}

{\it Experiment}
The experiment was performed under microgravity conditions in the Russian-German PK-3~Plus Laboratory on board the International Space Station (ISS). Technical details of the setup can be found in \cite{Thomas:2008}. An argon plasma was produced by a capacitively-coupled radio-frequency (rf) generator in push-pull mode at $13.56$~MHz. We prepared a binary complex plasma by injecting two types of particles. The first type is melamine formaldehyde (MF) particles of a diameter of $2.55$~$\mu$m with a mass $m_s=1.34\times10^{-14}$~kg, while the second type is SiO$_2$ particles of a diameter of $1.55$~$\mu$m with a mass $m_b=3.6\times10^{-15}$~kg. With video microscopy \cite{Thomas:2008}, a cross section of particle cloud \footnote{The video is recorded by the High Resolution camera, which records the upper-center portion of the whole particle cloud.} (illuminated by a laser sheet) was recorded with a frame rate of $50$~frames-per-second (fps) and a spatial resolution of $0.01$~mm/pixel. This makes it possible to trace the particles from frame to frame and calculate velocities and accelerations. The gas pressure was set at $10$~Pa, and the discharge voltage was set at $30$~V.

As we can see in Fig.~\ref{figure1}(a), the two particle types were phase-separated with a clear interface, mainly due to the difference of the ion drag force\cite{Killer:2016}. The big particles were confined in the upper part while the small particles were located in the lower part. The solitary wave was excited by switching on and off the function generator (FG) \cite{Schwabe:2008}. As the FG was switched off, the particle cloud recorded in the HR camera was pressed downwards. This left a trace as the red area on the top-left corner in Fig.~\ref{figure1}. As the cloud hit the edge of the central  void, the downwards motion was stopped and the quasi-solitary wave was excited. The propagation of this wave is marked in Fig.~\ref{figure1}(a-e). The wave front is  highlighted by the black frame. As we see in Fig.~\ref{figure1}(l), the solitary wave propagates with a phase speed of $25$~mm/s in the sub-cloud of small particles and $15$~mm/s in that of big particles.

{\it Simulations}
We complement the experiment with simulations in order to find out under what conditions the solitary wave is reflected at the interface. Langevin dynamics simulations are widely used to study the motion of microparticles in complex plasmas \cite{Hou:2009,Jiang:2011,Schwabe:2013}. The equation of motion including damping from the neutral gas and Brownian motion of microparticles is given by:
\begin{equation}
m_i\ddot{\bm{r}}_i+m_i{\nu_i}\dot{\bm{r}}_i=-\sum\limits_{j \ne i}{\nabla\phi_{ij}}+\bm{F}_{id,i}+\bm{F}_{c,i}+\bm{L}_i,
\label{eq1}
\end{equation}
where $\bm{r_i}$ is the three-dimensional particle position, $m_i$ the mass, $\nu_i$ the damping rate, $\bm{L}_i$ the Langevin heat bath. The Langevin force $\bm{L}_i$ is defined by $<\bm{L}_i(t)>=0$ and $< \bm{L}_i(t+\delta)\bm{L}_i(t+\delta)>=2{\nu_i}{m_i}T\delta(\tau)$, where $T$ is the temperature of the heat bath, $\delta(t)$ is the delta function.

The force acting on the particle $i$ includes three components. The first term on the right hand side of Eq.~(\ref{eq1})  is the sum of the Yukawa interaction with neighboring particles
\begin{equation}
\phi_{ij}=\frac{Q_iQ_j}{4\pi\epsilon_0r_{ij}}\exp(-\frac{r_{ij}}{\lambda}),
\end{equation}
where $\lambda$ is the Debye length, $Q_i$ is the charge of particle $i$ and $Q_j$ is the charge of a neighboring particle $j$ . The second term $\bm{F}_{id}$ is the ion drag force directed in the positive $z$ direction (upwards). We assume two constant ion drag forces for small and big particles. The third term is the plasma confinement force $\bm{F}_{c,i}=-\nabla{\Psi}Q_i$, where the confinement potential reads $\Psi=1/2Cz^2$, forming a parabolic confinement with a constant confinement coefficient $C$. The position of the particle cloud is determined by the plasma potential and the ion drag force. Due to the difference in the magnitude of the ion drag force and the particle charges, the particle cloud is phase separated where the small particles are located below the big particles, as we see in Fig.~\ref{figure1}. The  molecular-dynamics simulations were performed using LAMMPS in NVE ensemble \cite{Plimpton:1995,lammps}.

\begin{figure}[!ht]
\centerline{\includegraphics[bb=0 20 700 700, width=0.5\textwidth]{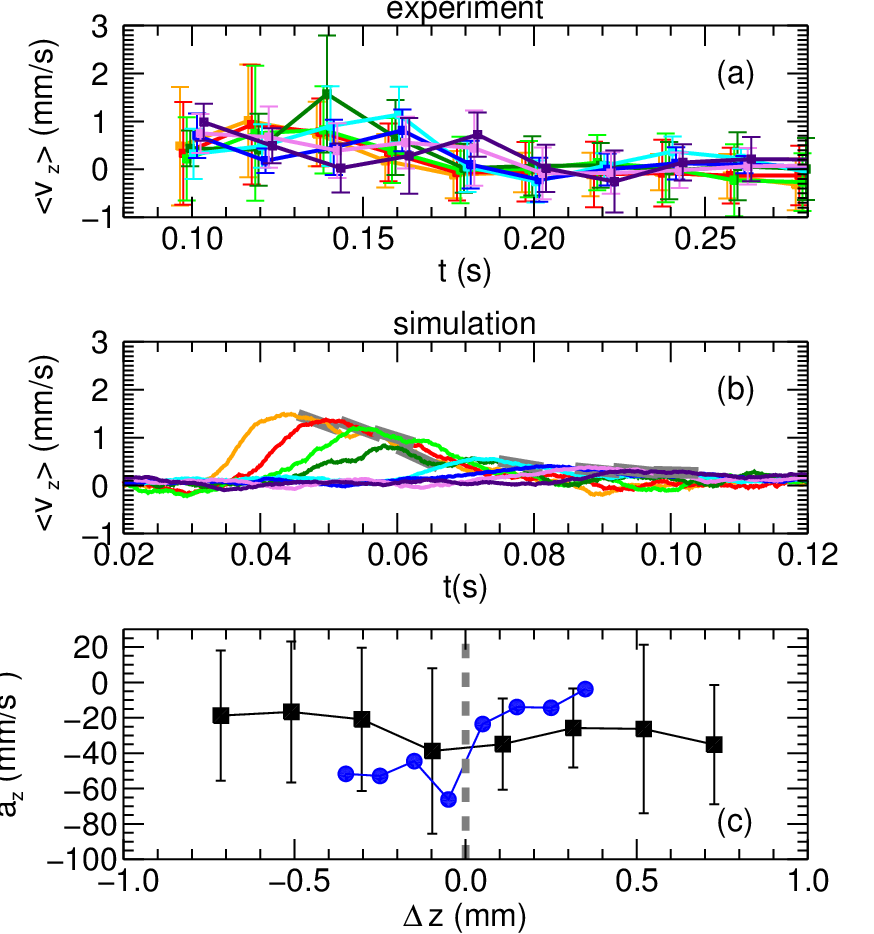}}
\caption{(Color online) Evolution of particle velocity at different heights in the experiment (a) and in the simulation (b). The curves colored from orange to green represent small particles located in front of the interface. The curves colored from cyan to black represent big particles located behind the interface. The acceleration of particles after the wave front (peak in the velocity evolution) is measured directly in the experiment and by a linear fit in the simulation, highlighted by a grey line. The dependence of the particle acceleration on the distance to the interface $\Delta z$ (c). The black squares correspond to the experiment, the blue circles to the simulation. The interface position is marked by the grey line.
\label{figure2}}
\end{figure}

The simulations were performed for $4000$ small particles and $4000$ big particles with periodic boundaries in $x$ and $y$ directions. We chose plasma and particle parameters according to typical parameters in experiments \cite{Yang:2017,Thomas:2008,Schwabe:2018}. Two types of particles were selected (the same as in the experiment). The charge for small and big particles was set as $Q_s=2700$~e and $Q_b=4500$~e, respectively. The damping rate was  assumed to be $\nu_s=57$~s$^{-1}$ and $\nu_b=42$~s$^{-1}$, corresponding to a gas pressure of $10$~Pa. The Debye length is $\lambda=100$~$\mu$m. The temperature of the Langevin heat bath is set as $T=300$~K for both particle types. The ion drag force for the small particles is set as $F_{id,s}=1$~fN while for big particles is $F_{id,b}=2.5$~fN, pushing the particles upwards. The confinement coefficient is $C=3000$~V/m$^2$.

As first step, the particles were allowed to move to an equilibrium position from their initial positions, where small particles were placed close to the piston with big particles above them (already phase separated). Then, the quasi-solitary wave was excited by a push from a piston, shown in Fig.~\ref{figure1}. The piston was moved upwards in a form of $z=A\sin(2{\pi}t/t_p)$ where the amplitude $A$ is $0.1$~mm and period is $t_p$ is $0.12$~s. The piston pushes for the first quarter period, namely $t_{push}=0.03$~s, and keeps still afterwards. While the excitation mechanism in the simulation is slightly different from that in the experiment, the resulting soliton propagation is comparable, as we shall see next.

The propagation of the solitary wave is shown in Fig.~\ref{figure1} and in the movies in the supplemental material. As the piston moves upwards, the particles close the piston were dramatically accelerated, forming the wave crest of the solitary wave. This solitary wave propagated upwards with a velocity of $25$~mm/s and encountered the interface. As we can see in Fig.~\ref{figure1}(f,g), the interface moved slightly upwards with a distance similar to the amplitude of the piston. The particle velocity in the soliton becomes much smaller after crossing the interface and further propagates upwards. The wave speed in the big particle cloud drops to $12$~mm/s.

\begin{figure}[!ht]
\centerline{\includegraphics[bb=0 20 700 450, width=0.5\textwidth]{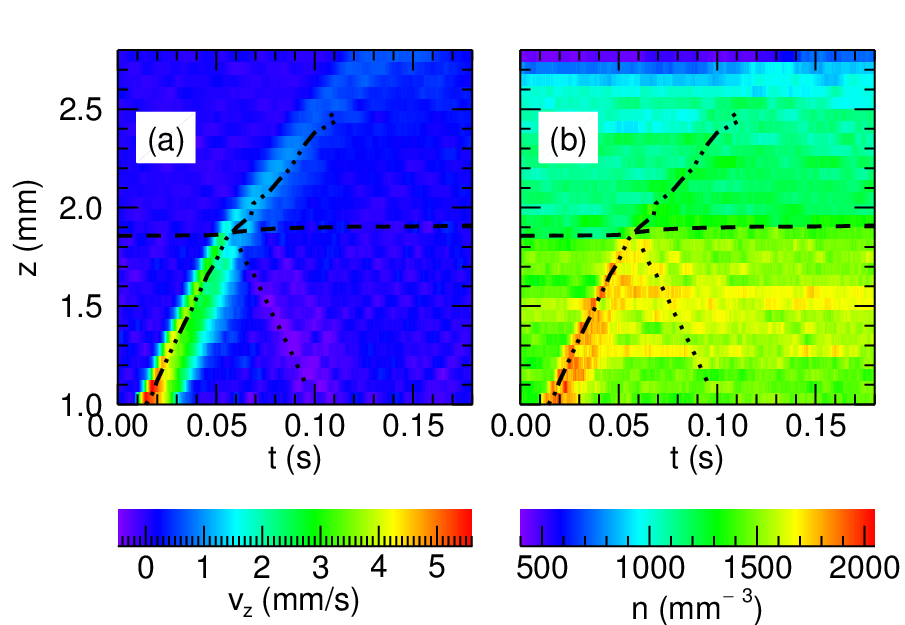}}
\caption{(Color online) The evolution of the simulated solitary wave by particle velocity (a) and number density (b) at a gas pressure of $p = 5$~Pa. The interface position is marked by a dashed line. The wave crest is highlighted by a dashed-dotted line, and the reflection is highlighted by the dotted line. See also the supplemental material.
\label{figure3}}
\end{figure}

{\it Interfacial effect}
The effect of the interface can be studied by measuring the particle deceleration behind the crest of the solitary wave. We divide the particle cloud into cells in $z$ direction and plot the evolution of $v_z$ in each cell. For the experiment, we calculated the velocity distribution and plot the mean value and the standard deviation as error in Fig.~\ref{figure2}(a). The deceleration for each height was directly measured and is shown as black square with error bars in Fig.~\ref{figure2}(c). As we can see, the deceleration becomes stronger (from $-20$~mm/s$^2$ to $-40$~mm/s$^2$) right before the interface. As the wave propagates in the sub-cloud of big particles, the deceleration is restored to the level of about $-20$~mm/s$^2$. This drop shows the effect of interface.

In the case of the simulation, the particle deceleration behind the wave crest is measured by a linear fit to the velocity curve behind the peak, as marked by the grey shadows in Fig.~\ref{figure2}(b). As we can see in the figure, the deceleration keeps almost constant at $-50$~mm/s$^2$ in the sub-cloud of small particles but drops dramatically to $-70$~mm/s$^2$ close to the interface, signifying an interface effect. The deceleration becomes much weaker as the wave passes the interface. The value reaches $-20$~mm/s$^2$ while the solitary wave propagates in the sub-cloud of big particles. The simulation shows a quantitative agreement with the experiment. The subtle difference in the magnitude may be caused by the simplification of the configuration of the confinement and the strength of the ion drag force.

{\it Reflection} In both experiment and simulation, the solitary waves decay significantly as they  propagate. This is mainly caused by the neutral gas friction. Though the gas pressure can not be further decreased in the experiment in PK-3 Plus, we can reduce the friction in the simulation to study the subtle feature of the interaction between the soliton and the interface. In the remainder of this paper, we set the gas pressure to $5$ Pa and moderately increase the strength of the push in the simulation with $A=1.5$~mm and $t_p=0.12$~s.

As we can see in Fig.~\ref{figure3}(a), for the lower pressure and stronger push the particle velocity in the crest of the solitary wave (highlighted by the dotted dashed line) increases dramatically compared to the high pressure case. Despite of the drop of the velocity at the interface, the propagation of the solitary wave is still clearly visible in the sub-cloud of the big particles. We also can see the propagation of the solitary wave in the density map in Fig.~\ref{figure3}(b). Remarkably, the signal of reflection emerges as a purple trace with negative particle velocity. We mark this trace with a dotted line.

\begin{figure}[!t]
\centerline{\includegraphics[bb= 0 0 700 700, width=0.5\textwidth]{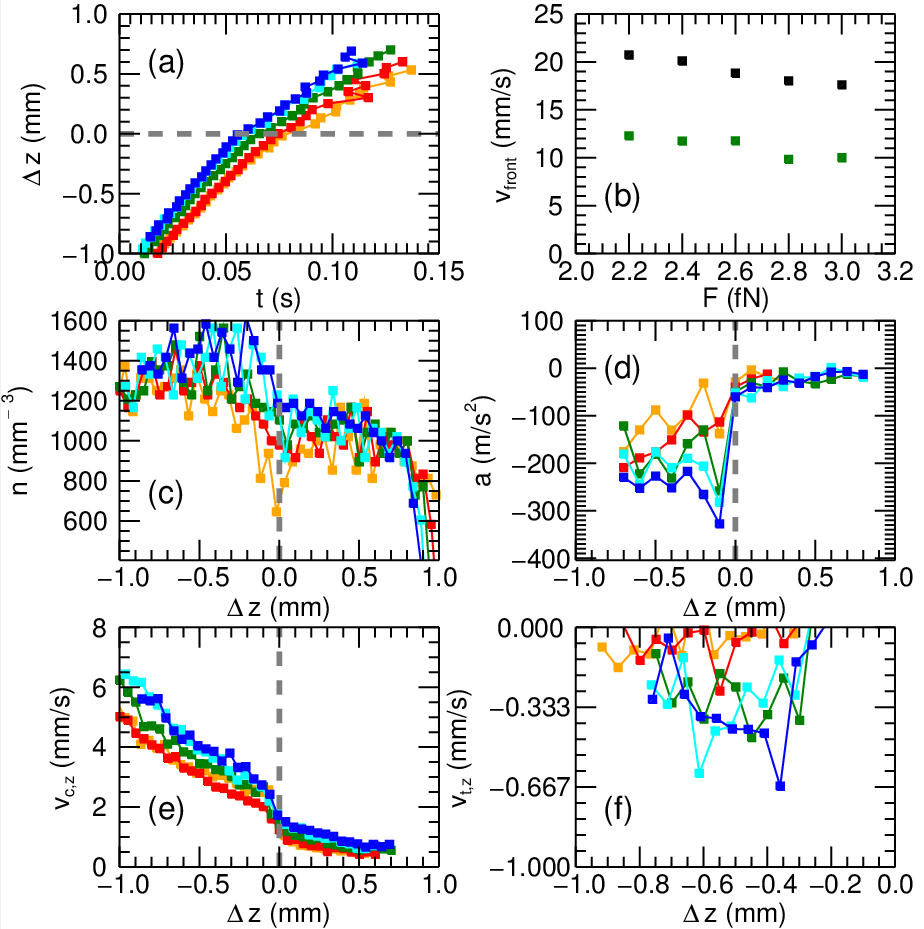}}
\caption{(Color online) Dependence of the evolution of the quasi-solitary wave on the applied ion drag force exerted on the big particles (a). Five ion drag forces are probed: orange for $2.2$~fN, red for $2.4$~fN, green for $2.6$~fN, cyan for $2.8$~fN, and blue for $3.0$~fN. The speed of the solitary wave in the clouds composed of small (black) and big (green) particles (b). Other properties including the density (c), acceleration of particles after the wave front (d), peak particle velocity in the wave front (e), and particle velocity in the reflection(f).
\label{figure4}}
\end{figure}

{\it Dependence on the width of the interface} In order to study the relation of the interfacial effect to the interface width, we tune the ion drag acting on the big particles as a control parameter. For better comparison, we define a relative distance $\Delta z=z-z_0$ where $z_0$ is the position of the interface (marked by grey dashed line), and focus on the range close to it, namely $-1$~mm~$<\Delta z <1$~mm. By increasing $F_{id,b}$, the big particle sub-cloud is dragged slightly upwards. The number density of the small particles decreases dramatically. As we can see in Fig.~\ref{figure4}(c), for $F_{id,b}=3$~fN, the small particle number density decreases from $1200$~mm$^{-3}$ close to the piston to $800$~mm$^{-3}$ at the interface. This number density keeps stable in the big particle cloud. On the other hand, for small ion drag force, the small particle cloud is compressed into a smaller area by the big particle cloud, resulting in a higher number density. As we can see in Fig.~\ref{figure4}(c), the density in the small particle sub-cloud first increases to $1600$~mm$^{-3}$ and then decreases to $1200$~mm$^{-3}$.

Decreasing the ion drag force also results in a narrower interface. In fact, as we set $F_{id,b}$ to $3.0$, $2.8$, $2.6$, $2.4$ and $2.2$~fN, the width of interface\footnote{The width is defined as the vertical distance between the average height of the $100$ lowest big particles and the $100$ highest small particles, to suppress the outliers.} is $0.005$, $0.02$, $0.03$, $0.05$, and $0.07$~mm, respectively.

However, as we can see in Fig.~\ref{figure4}(a,b), the speed of the solitary wave does not depend strongly on the particle number density and the width of the interface. The speed of the soliton in the small particle cloud decreases slightly from $21$~mm/s to $17$~mm/s as the ion drag force increases. Similarly, the speed in the big particle cloud decrease from $13$~mm/s to $10$~mm/s.

The interfacial effect, illustrated by the particle deceleration behind the wave front, depends strongly on the width of the interface. As the solitary wave propagates in the sub-cloud of small particles, deceleration of particles in the wave crest increases dramatically from $-100$~mm/s to $-250$~mm/s as the ion drag force on the big particles increases, as we see in Fig.~\ref{figure4}(d). This is mainly caused by the decrease of the number density of the small particles. In the case of the big particles, the magnitude of the deceleration is comparable for all values of the ion drag force due to similar number density. As to the interfacial effect, namely the spike of the deceleration close to the interface, the larger the interface is, the weaker the interfacial effect is. A smaller interface  causes a harder transition  that the small particles encounter. This leads to the higher deceleration.

Finally, we look into the maximal (in magnitude) particle velocity of the reflected solitary wave, as shown in Fig.~\ref{figure4}(f). As we can see, due to the strong interfacial effect and high deceleration, the particle velocity in the reflection is higher for the run with larger interface width though it decreases fast to the thermal velocity so that the reflection is barely noticeable. For larger interface width, the maximal velocity is lower. However, the overall difference is not very big.

{\it Conclusion}
In summary, we study the propagation of a dissipative solitary wave across an interface in a binary complex plasma. The experiments were performed in the PK-3 Plus Laboratory on board the International Space Station. The solitary wave was excited at the edge of the void and propagated from the sub-cloud of small particles into the that of big particles. The interfacial effect was observed by measuring the deceleration of particles behind the wave crest. The results are compared with a Langevin dynamics simulation, where the wave was excited by a gentle push on the edge of the sub-cloud of small particles. The interfacial effect shows a qualitative agreement between the experiments and the simulation. By increasing the strength of the push, reflection is observed. By tuning the ion drag force exerted on the big particles in the simulation, the width of the interface is adjusted. We found that the smaller the interface is, the stronger the interfacial effect is. The strength of the reflection is positively correlated to the magnitude of the interfacial effect.

\begin{acknowledgments}
The authors acknowledge support from the National Natural Science Foundation of China (NSFC), Grant No. 11405030. The PK-3 Plus project was funded by the space agency of the Deutsches Zentrum f\"ur Luft- und Raumfahrt eV with funds from the Federal Ministry for Economy and Technology according to a resolution of the Deutscher Bundestag under grant number 50 WM 1203. The authors acknowledge Roscosmos provided the PK-3 Plus laboratory launch and operation onboard of ISS. We thank Milenko Rubin-Zuzic for helpful comments.
\end{acknowledgments}

\end{document}